\documentclass{article}
\usepackage{amssymb}

\usepackage{sw20aip}

\input{tcilatex}
\begin{document}

\author{Ling-Fong Li, \and Physics Department, Carnegie Mellon University, \and %
Pittsburgh, PA 15213, USA}
\title{Spontaneous Symmetry Breaking and Chiral Symmetry\thanks{%
Lecture given at VII Mexico Workshop on Particles and Fields, Merida,
Yucatan Mexico, Nov 10-17 1999}}
\date{}
\maketitle

\begin{abstract}
In this introductory lecture, some basic features of the spontaneous
symmetry breaking are discussed. More specifically, $\sigma $-model,
non-linear realization, and some examples of spontaneous symmetry breaking
in the non-relativistic system are discussed in details. The approach here
is more pedagogical than rigorous and the purpose is to get some simple
explanation of some useful topics in this rather wide area. .
\end{abstract}

\section{Introduction}

The symmetry principle is perhaps the most important ingredient in the
development of high energy physics. Roughly speaking, the symmetries of the
physical system lead to conservation laws, which give many important
relations among physical processes. Many of the symmetries in Nature are
however approximate symmetries rather than exact symmetries and are also
very useful in the understanding of various phenomena in high energy
physics. Among the broken approximate symmetries, the most interesting one
is the Spontaneous Symmetry Breaking (SSB) which seems to have played a
special role in high energy physics. Many important progress has come from
the understanding of the SSB. The SSB is characterized by the fact that
symmetry breaking shows up in the ground state rather than in the basic
interaction. This makes it difficult to uncover this kind of approximate
symmetries. Historically, SSB was first discovered around 1960 in the study
of superconductivity in the solid state physics by Nambu and Goldstone \cite
{nambu,goldstone}. One of consequences of SSB is the presence of the
massless excitation \cite{salamwei}, called the Nambu-Goldstone boson, or
just Goldstone boson for short. Later, Nambu \cite{nambu} applied the idea
to the particle physics. In combination with $SU(3)\times SU(3)$ current
algebra, SSB has been quite successful in the understanding of the chiral
symmetry in the low energy phenomenology of strong interaction. More
importantly, in 1964 it was discovered by Higgs \cite{higgs} and others \cite
{englert,Guralnik} that in the context of gauge theory, SSB has the
remarkable property that it can convert the long range force in the gauge
theory into a short range force. Thus it avoids both the massless Goldstone
bosons and the massless gauge bosons. Weinberg, \cite{weinberg} and Salam 
\cite{salam}, then applied this ideas to construct a model of
electromagnetic and weak interactions. The significance of this model was
not realized until t'Hooft \cite{thooft} show in 1971 that it was
renomalizable. Since then this model has enjoyed remarkable experimental
success and now called the ``Standard Model of Electroweak Interactions'' 
\cite{standmod}. Undoubtedly, this will serve as benchmark for any new
physics for years to come.

In this article I will give a simple introduction to the spontaneous
symmetry breaking and its application to chiral symmetries in the hadronic
interaction. The emphasis is on the qualitative understanding rather than
completeness and mathematical rigor. Eventhough SSB has been quite
successful in explaining many interesting phenomena, its implementation in
the theoretical framework is more or less put in by hand and it is not at
all clear what is the origin of SSB. Here I will also discuss some
non-relativistic example where the physics is more tractable in the hope
that they might give some hints about the true nature of SSB. Maybe good
understanding of SSB might extend its applicability to some new frontier.

\section{$\protect\smallskip $\protect\underline{$SU\left( 2\right) \times
SU\left( 2\right) $ $\sigma $-Model}}

The $\sigma $-model has a long and interesting history. It was originally
constructed in 1960's as a tool to study the chiral symmetry in the system
with pions and nucleons\cite{Gellmann}. Later the spontaneous symmetry
breaking and PCAC (partially conserved axial current) were incorporated..
Eventhough this model is not quite phenomenologically correct it remains the
simplest example which realizes many important aspects of broken symmetries.
Even though the strong interaction is now described by $QCD,$ the $\sigma $%
-model of pions and nucleons is still useful as an effective interaction in
the low energies where it is difficult to calculate directly from $QCD.$ In
addition, the $\sigma $-model has also been used quite often as a framework
to test many interesting ideas in field theory and string theory. Here we
will discuss the most basic features of the $\sigma $-model.

The Lagrangian for $SU\left( 2\right) \times SU\left( 2\right) $ $\sigma $%
-Model is given by 
\begin{eqnarray}
\mathcal{L} &=&\frac{1}{2}\left[ \left( \partial _{\mu }\sigma \right)
^{2}+\left( \partial _{\mu }\stackrel{\rightarrow }{\pi }\right) ^{2}\right]
+\frac{\mu ^{2}}{2}\left( \sigma ^{2}+\stackrel{\rightarrow }{\pi }%
^{2}\right) -\frac{\lambda }{4}\left( \sigma ^{2}+\stackrel{\rightarrow }{%
\pi }^{2}\right) ^{2}  \label{sigma} \\
&&+\overline{N}i\gamma ^{\mu }\partial _{\mu }N+g\overline{N}\left( \sigma
+i\gamma _{5}\stackrel{\rightarrow }{\tau }\cdot \stackrel{\rightarrow }{\pi 
}\right) N  \nonumber
\end{eqnarray}
where $\stackrel{\rightarrow }{\pi }=\left( \pi _{1},\pi _{2},\pi
_{3}\right) $ is the isotriplet pion fields, $\sigma $ is the isosinglet
field, and $N$ is the isodoublet nucleon field. To discuss the symmetry
property, it is more useful to write this Lagrangian as :\newline
\begin{eqnarray}
\mathcal{L} &=&\frac{1}{2}tr\left( \partial _{\mu }\Sigma \partial ^{\mu
}\Sigma ^{\dagger }\right) +\frac{\mu ^{2}}{4}tr\left( \Sigma \Sigma
^{\dagger }\right) -\frac{\lambda }{8}\left[ \left( \Sigma \Sigma ^{\dagger
}\right) \right] ^{2}  \label{sigma2} \\
&&+\overline{N}_{L}i\gamma ^{\mu }\partial _{\mu }N_{L}+\overline{N}%
_{R}i\gamma ^{\mu }\partial _{\mu }N_{R}+g(\overline{N}_{L}\Sigma N_{R}+%
\overline{N}_{R}\Sigma ^{\dagger }N_{L})  \nonumber
\end{eqnarray}
where 
\begin{equation}
\Sigma =\sigma +i\stackrel{\rightarrow }{\tau }\cdot \stackrel{\rightarrow }{%
\pi },\qquad N_{L}=\frac{1}{2}\left( 1-\gamma _{5}\right) N,\qquad N_{R}=%
\frac{1}{2}\left( 1+\gamma _{5}\right) N
\end{equation}
This Lagrangian is now clearly invariant under transformation, 
\begin{equation}
\Sigma \rightarrow \Sigma ^{\prime }=L\Sigma R^{\dagger },\qquad
N_{L}\rightarrow N_{L}^{\prime }=LN_{L},\qquad N_{R}\rightarrow
N_{R}^{\prime }=RN_{R}
\end{equation}
where 
\begin{equation}
L=\exp \left( -i\stackrel{\rightarrow }{\tau }\cdot \stackrel{\rightarrow }{%
\theta }_{L}\right) ,\qquad R=\exp \left( -i\stackrel{\rightarrow }{\tau }%
\cdot \stackrel{\rightarrow }{\theta }_{R}\right)
\end{equation}
are two arbitrary $2\times 2$unitary matrices. Thus the symmetry group is $%
SU\left( 2\right) _{L}\times SU\left( 2\right) _{R}$ and representation
contents under this group are 
\[
\Sigma \sim \left( \frac{1}{2},\frac{1}{2}\right) ,\qquad N_{L}\sim \left( 
\frac{1}{2},0\right) ,\qquad N_{R}\sim \left( 0,\frac{1}{2}\right) 
\]
\underline{Remark}:The nucleon mass term $\overline{N}_{L}N_{R}+h.c.$%
transforms as $\left( \frac{1}{2},\frac{1}{2}\right) $ representation and is
not invariant. One way to construct invariant nucleon mass term is to
introduce another doublet of fermions with opposite parity, 
\[
N_{L}^{\prime }\sim \left( 0,\frac{1}{2}\right) ,\qquad N_{R}^{\prime }\sim
\left( \frac{1}{2},0\right) 
\]
so that the term $\left( \overline{N}_{L}^{\prime }N_{R}+\overline{N}%
_{R}^{\prime }N_{L}+h.c.\right) $is invariant. This will give same mass to
both doublets and is usually called parity doubling. As we shall see later,
another way to give mass to nucleon is by spontaneous symmetry breaking
which does not require another doublet.

The general form of Noether current is of the form 
\[
J_{\mu }\sim \stackunder{i}{\sum }\frac{\partial \mathcal{L}}{\partial
\left( \partial _{\mu }\phi _{i}\right) }\delta \phi _{i} 
\]
where $\delta \phi _{i}$ is the infinitesimal change of the fields under the
symmetry transformations. We have for the left-handed transformation, 
\begin{eqnarray}
\delta _{L}\sigma &=&\stackrel{\rightarrow }{\theta }_{L}\cdot \stackrel{%
\rightarrow }{\pi },\qquad \delta _{L}\stackrel{\rightarrow }{\pi }=-%
\stackrel{\rightarrow }{\theta }_{L}\sigma +\stackrel{\rightarrow }{\theta }%
_{L}\times \stackrel{\rightarrow }{\pi },  \label{lefttran} \\
\qquad \delta _{L}N_{L} &=&-i\frac{\stackrel{\rightarrow }{\theta }_{L}\cdot 
\stackrel{\rightarrow }{\tau }}{2}N_{L},\qquad \delta _{L}N_{R}=0  \nonumber
\end{eqnarray}
and 
\begin{equation}
J_{L\mu }^{a}=\varepsilon ^{abc}\pi ^{b}\partial _{\mu }\pi ^{c}+\left[
\sigma \partial _{\mu }\pi ^{a}-\pi ^{a}\partial _{\mu }\sigma \right] +%
\overline{N}_{L}\gamma _{\mu }\frac{\tau ^{a}}{2}N_{L}
\end{equation}
Similarly, 
\begin{eqnarray}
\delta _{R}\sigma &=&-\stackrel{\rightarrow }{\theta }_{R}\cdot \stackrel{%
\rightarrow }{\pi },\qquad \delta _{R}\stackrel{\rightarrow }{\pi }=%
\stackrel{\rightarrow }{\theta }_{R}\sigma +\stackrel{\rightarrow }{\theta }%
_{R}\times \stackrel{\rightarrow }{\pi },  \label{righttran} \\
\qquad \delta _{R}N_{L} &=&0,\qquad \delta _{R}N_{R}=-i\frac{\stackrel{%
\rightarrow }{\theta }_{R}\cdot \stackrel{\rightarrow }{\tau }}{2}N_{R} 
\nonumber
\end{eqnarray}
and 
\begin{equation}
J_{R\mu }^{a}=\varepsilon ^{abc}\pi ^{b}\partial _{\mu }\pi ^{c}-\left[
\sigma \partial _{\mu }\pi ^{a}-\pi ^{a}\partial _{\mu }\sigma \right] +%
\overline{N}_{R}\gamma _{\mu }\frac{\tau ^{a}}{2}N_{R}
\end{equation}
The corresponding charges are given by 
\begin{equation}
Q_{L}^{a}=\int d^{3}xJ_{L0}^{a},\qquad Q_{R}^{a}=\int d^{3}xJ_{R0}^{a}.
\end{equation}
Using the canonical commutation relations, we can derive 
\begin{equation}
\left[ Q_{L}^{a},Q_{L}^{b}\right] =i\varepsilon _{ijk}Q_{L}^{c},\qquad
\left[ Q_{R}^{a},Q_{R}^{b}\right] =i\varepsilon _{ijk}Q_{R}^{c},\qquad
\left[ Q_{L}^{a},Q_{L}^{b}\right] =0
\end{equation}
which is the $SU_{L}\left( 2\right) \times SU_{R}\left( 2\right) $ algebra.

The vector and axial charges are given by 
\begin{equation}
Q^{a}=Q_{R}^{a}+Q_{L}^{a},\qquad Q_{5}^{a}=Q_{R}^{a}-Q_{L}^{a}
\end{equation}
In particular,the axial charges are 
\begin{equation}
Q_{i}^{5}=\int d^{3}xA_{i}^{0}\left( x\right) =\int d^{3}x\left[ i\left(
\sigma \partial _{0}\pi _{i}-\pi _{i}\partial _{0}\sigma \right) +N^{\dagger
}\frac{\sigma _{i}}{2}\gamma _{5}N\right]  \label{axialcurr}
\end{equation}
\underline{Remark}: Another way to describe the symmetry of the $\sigma $%
-model is the $O\left( 4\right) $ symmetry, which is isomorphic to $SU\left(
2\right) \times SU\left( 2\right) $ locally and is characterized by $4\times
4$ orthogonal matrix, 
\[
RR^{T}=R^{T}R=1 
\]
The infinitesimal transformation is 
\[
R_{ij}=\delta _{ij}+\varepsilon _{ij},\qquad \text{with\qquad }\varepsilon
_{ij}=-\varepsilon _{ji} 
\]
The scalar field $\phi _{i}=\left( \pi _{1},\pi _{2},\pi _{3},\sigma \right) 
$ transform as 4-dimensional vector, 
\[
\phi _{i}\rightarrow \phi _{i}^{\prime }=R_{ij}\phi _{j}\simeq \phi
_{i}+\varepsilon _{ij}\phi _{j} 
\]
The combination $\phi _{i}\phi _{i}=\sigma ^{2}+\stackrel{\rightarrow }{\pi }%
^{2}$is just the length of the vector $\phi _{i}$ and is clearly invariant
under the rotations in 4-dimension. If we take 
\[
\varepsilon _{ij}=\varepsilon _{ijk}\alpha _{k},\qquad \varepsilon
_{4i}=\beta _{i},\qquad i,j,k=1,2,3 
\]
we get 
\[
\stackrel{\rightarrow }{\pi }^{\prime }=\stackrel{\rightarrow }{\pi }+%
\stackrel{\rightarrow }{\alpha }\times \stackrel{\rightarrow }{\pi }+%
\stackrel{\rightarrow }{\beta }\sigma ,\qquad \sigma ^{\prime }=\sigma +%
\stackrel{\rightarrow }{\beta }\cdot \stackrel{\rightarrow }{\pi } 
\]
Thus we see from Eqs(\ref{lefttran},\ref{righttran}) that the parameters $%
\stackrel{\rightarrow }{\alpha }$ correspond to vector transformations and $%
\stackrel{\rightarrow }{\beta }$ the axial transformation.

\subsection{Spontaneous Symmetry Breaking}

The classical ground state is determined by minimum of the self interaction
of scalars, 
\begin{equation}
V\left( \sigma ,\stackrel{\rightarrow }{\pi }\right) =-\frac{\mu ^{2}}{2}%
\left( \sigma ^{2}+\stackrel{\rightarrow }{\pi }^{2}\right) +\frac{\lambda }{%
4}\left( \sigma ^{2}+\stackrel{\rightarrow }{\pi }^{2}\right) ^{2}
\end{equation}
The minimum of the potential is located at 
\begin{equation}
\sigma ^{2}+\stackrel{\rightarrow }{\pi }^{2}=\frac{\mu ^{2}}{\lambda }%
\equiv v^{2}
\end{equation}
which is a 3-sphere, $S^{3}$ in the 4-dimensional space formed by the scalar
fields. Each point on $S^{3}$ is invariant under $O\left( 3\right) $
rotations. For example, the point $\left( 0,0,0,v\right) $ is invariant
under the rotations of the first 3 components of the vector. Then, after a
point on $S^{3}$ is chosen to be the classical ground state, the symmetry is
broken spontaneously from $O\left( 4\right) $ to $O\left( 3\right) .$ Note
that different points on $S^{3}$ are related to each other by the action of
those rotations which are in $O\left( 4\right) $ but not in $O\left(
3\right) .$ These rotations are usually denoted by $O\left( 4\right)
/O\left( 3\right) .$(This is called the coset space.). Thus we can identify
3-sphere with $O\left( 4\right) /O\left( 3\right) .$

For the quantum theory, we need to expand the fields around the classical
values 
\begin{equation}
\sigma =v+\sigma ^{\prime },\quad \stackrel{\rightarrow }{\pi }^{\prime }=%
\stackrel{\rightarrow }{\pi },\quad \text{where }<\sigma >=v  \label{vev}
\end{equation}
Here $v$ is usually called the vacuum expectation value (VEV). Then we see
that 
\begin{equation}
V\left( \sigma ,\stackrel{\rightarrow }{\pi }\right) =\mu ^{2}\sigma
^{\prime 2}+\lambda v\sigma ^{\prime }\left( \sigma ^{\prime 2}+\stackrel{%
\rightarrow }{\pi ^{\prime }}^{2}\right) +\frac{\lambda }{4}\left( \sigma
^{\prime 2}+\stackrel{\rightarrow }{\pi ^{\prime }}^{2}\right) ^{2}
\label{ssbv}
\end{equation}
\[
g\overline{N}\left( \sigma +i\gamma _{5}\stackrel{\rightarrow }{\tau }\cdot 
\stackrel{\rightarrow }{\pi }\right) N=gv\overline{N}N+g\overline{N}\left(
\sigma ^{\prime }+i\gamma _{5}\stackrel{\rightarrow }{\tau }\cdot \stackrel{%
\rightarrow }{\pi ^{\prime }}\right) N 
\]
Thus $\pi ^{\prime }s$ are massless and $N$'s are massive.\newline
Remark:If we had made another choice for VEV,e.g. 
\[
<\pi _{3}>=v,\qquad <\pi _{1}>=<\pi _{2}>=<\sigma >=0 
\]
The physics is still the same as we will now illustrate. In this case, we
write $\pi _{3}^{^{\prime }}=\pi _{3}+v$ to get 
\[
V\left( \sigma ,\stackrel{\rightarrow }{\pi }\right) =\mu ^{2}\pi ^{\prime
2}+\left( \text{cubic terms and higher}\right) 
\]
\[
g\overline{N}\left( \sigma +i\gamma _{5}\stackrel{\rightarrow }{\tau }\cdot 
\stackrel{\rightarrow }{\pi }\right) N=gv\overline{N}\tau _{3}\gamma
_{5}N+\cdots 
\]
Thus we still have 3 massless scalar fields. For the nucleon, if we define 
\[
N_{L}=\exp \left( -i\pi \frac{\tau _{3}}{2}\right) N_{L}^{^{\prime }},\qquad
N_{R}=N_{R}^{^{\prime }} 
\]
we get 
\[
gv\overline{N}\tau _{3}\gamma _{5}N=gv\overline{N}^{^{\prime }}N^{^{\prime
}} 
\]
which the mass term for the new field. It is easy to see that $%
Q_{5}^{1},Q_{5}^{2},Q$ $^{3}$form an unbroken $SU\left( 2\right) $ algebra.

There are several interesting features worth noting:

$\left( 1\right) \stackrel{\rightarrow }{\pi }$'s are massless. This is a
consequence of the Goldstone theorem which states that spontaneous symmetry
breaking (SSB)of a continuous symmetry will give massless particle or zero
energy excitation. This theorem will be discussed in more detail in next
subsection.

$\left( 2\right) $After SSB, the original multiplet $\left( \sigma ,%
\stackrel{\rightarrow }{\pi }\right) $splits into massless $\stackrel{%
\rightarrow }{\pi }^{\prime }s$ and massive $\sigma .$ Also the nucleons
become massive. Thus eventhough the interaction is $SU\left( 2\right)
_{L}\times SU\left( 2\right) _{R}$ symmetric the spectrum is only $SU\left(
2\right) $ symmetric. This is the typical consequence of SSB. In some sense,
the original symmetry is realized by combining the $SU\left( 2\right) $
multiplet, e.g. $N$ ,with the massless Goldstone bosons to form the
multiplets of $SU\left( 2\right) _{L}\times SU\left( 2\right) _{R}.$ This,
as we will discuss later, is the basis of the low energy theorem.

$\left( 3\right) $The axial current in Eq(\ref{axialcurr}) after the SSB
will have a term linear in $\pi $ field, 
\begin{equation}
A_{i}^{\mu }=iv\partial _{\mu }\pi _{i}+\cdots
\end{equation}
which is responsible for the matrix element, 
\begin{equation}
<0|A_{i}^{\mu }|\pi _{j}\left( p\right) >=ip^{\mu }v
\end{equation}
Using this matrix element in $\pi $ decay, we can identify the VEV $v$ with
the pion decay constant $f_{\pi }.$ This coupling between axial current $%
A_{i}^{\mu }$ and $\pi _{i}$ will give rise to a massless pole.

$\left( 4\right) $The appearance of the cubic term $\sigma ^{^{\prime
}}\left( \sigma ^{\prime 2}+\stackrel{\rightarrow }{\pi ^{\prime }}%
^{2}\right) $ and the mass term $gv\overline{N}N$ is the result of the
spontaneous symmetry breaking . Since these terms have dimension 3, they are
usually called soft breaking, in contrast to the dimension 4 hard breaking
terms.

$\left( 5\right) $In the scalar self interaction, quartic, cubic, and
quadratic terms have only 2 parameters, $\lambda ,\mu .$ This means that
these 3 terms are not independent, and there is a relation among them. This
is an example of low energy theorem for theory with spontaneous symmetry
breaking.

\subsection{Low energy theorem}

The SSB leads to many relations which are quite different from the usual
symmetry breaking. The most distinct ones are relations among amplitudes
involving Goldstone bosons in low energies. As we have mentioned before,
these relations are consequence of the fact that Goldstone bosons are
massless and can be tagged on to other particles to form a larger multiplet.
Since Goldstone bosons do carry energies, this is possible only in limit
that Goldstone bosons have zero energies.

Consider the following processes involving the Goldstone bosons in the
external states.

$\left( i\right) $\underline{$\pi ^{0}\left( p_{1}\right) +\sigma \left(
p_{2}\right) \rightarrow \pi ^{0}\left( p_{3}\right) +\sigma \left(
p_{4}\right) $}\newline
The tree-level contributions are coming from diagrams in Fig1,\bigskip 
\FRAME{dtbpFU}{13.0919cm}{5.8957cm}{0pt}{\Qcb{Fig 1 Tree graphs for $\pi \pi 
$ scattering}}{\Qlb{Fig1}}{fig1.wmf}{\special{language "Scientific
Word";type "GRAPHIC";display "USEDEF";valid_file "F";width 13.0919cm;height
5.8957cm;depth 0pt;original-width 723.0625pt;original-height
542.0625pt;cropleft "-0.0035606";croptop "0.9848812";cropright
"0.9349394";cropbottom "0.0440812";filename 'A:/fig1.wmf';file-properties
"XNPEU";}}

The amplitudes for these diagrams are given by, 
\begin{equation}
M_{a}=\left( -2i\lambda v\right) ^{2}\frac{i}{s},\quad M_{b}=3\left(
-2i\lambda v\right) ^{2}\frac{i}{t-m_{\sigma }^{2}}.\qquad M_{c}=\left(
-2i\lambda v\right) ^{2}\frac{i}{u},\quad M_{d}=-2i\lambda
\end{equation}
\begin{equation}
M=M_{a}+M_{b}+M_{c}+M_{d}=4i\lambda ^{2}v^{2}\left[ \frac{1}{s}+\frac{3}{%
t-m_{\sigma }^{2}}+\frac{1}{u}+\frac{1}{2\lambda v^{2}}\right]
\end{equation}
Here $s,$ $t,$ and $u$ are the usual Mandelstam variables, $s=\left(
p_{1}+p_{2}\right) ^{2},t=\left( p_{1}-p_{3}\right) ^{2},u=\left(
p_{1}-p_{4}\right) ^{2}$. In the limit where pions have zero momenta, $%
p_{1}=p_{3}=0,$ we get $s=u=m_{\sigma }^{2},$ $t=0$ and 
\begin{equation}
M=4i\lambda ^{2}v^{2}\left[ \frac{1}{m_{\sigma }^{2}}+\frac{1}{m_{\sigma
}^{2}}-\frac{3}{m_{\sigma }^{2}}+\frac{1}{m_{\sigma }^{2}}\right] =0
\end{equation}
where we have used $m_{\sigma }^{2}=2\lambda \upsilon ^{2}.$ Thus the
amplitude vanishes in the soft pion limit, i.e. $p_{\pi }\rightarrow
0.\bigskip $

$\left( ii\right) $\underline{$\pi ^{0}\pi ^{0}\rightarrow \pi ^{0}\pi ^{0}$}
\newline
Similar calculation gives

\[
M=M_{a}+M_{b}+M_{c}+M_{4}=-2i\lambda \left[ \frac{s}{s-m_{\sigma }^{2}}+%
\frac{t}{t-m_{\sigma }^{2}}+\frac{u}{u-m_{\sigma }^{2}}\right] 
\]
In the soft pion limit, $p_{i}\rightarrow 0,$ we get 
\begin{equation}
M\simeq \frac{2i\lambda }{m_{\sigma }^{2}}\left( s+t+u\right) =\frac{i}{v^{2}%
}\left( s+t+u\right) \rightarrow 0.  \label{pipi}
\end{equation}
This is the same as the limit, $m_{\sigma }^{2}\rightarrow \infty ,$ because
soft pion means pion momentum much smaller than $m_{\sigma }^{2}.$ These are
simple examples of the low energy theorem which says that physical
amplitudes vanish in the limit where myomata of Goldstone bosons go to zero.

In examples above, the vanishing of these amplitudes results from some
cancellation among different contributions. Since this is a general property
of the Goldstone boson, there should be a better way of getting this. It
turns that one can change the variables representing the scalar fields such
that Goldstone bosons always enter with derivative coupling. Then the
vanishing of the amplitudes involving Goldstone boson is manifest. This can
be accomplished by the \underline{field redefinition} which we will now
describe briefly(\cite{haag}). Suppose we start from a Lagrangian with field 
$\phi $ and make a transformation to a new field $\eta ,$ with the relation, 
\[
\phi =\eta F\left( \eta \right) 
\]
where $F\left( \eta \right) $is some power series in $\eta .$If we impose
the condition that $F\left( 0\right) =1,$ the free Lagrangian for $\eta $
will be the same as that for $\phi .$ Then according to a general theorem
valid with rather weak restrictions on the Lagrangian and $F\left( \eta
\right) ,$ the on-shell matrix elements calculated with $\eta $ fields and
with $\phi $ fields are the same. We will use this field redefinition to
write the Goldstone boson interaction in terms of derivative coupling.
Consider a simplified Lagrangian given by

\begin{equation}
\mathcal{L}=\frac{1}{2}\left[ \left( \partial _{\mu }\sigma \right)
^{2}+\left( \partial _{\mu }\pi \right) ^{2}\right] +\frac{\mu ^{2}}{2}%
\left( \sigma ^{2}+\pi ^{2}\right) -\frac{\lambda }{4}\left( \sigma ^{2}+\pi
^{2}\right) ^{2}
\end{equation}
which is just the $O\left( 2\right) $ version of the $\sigma $-model without
the nucleon. As before , the SSB will require the shift of the $\sigma $%
-field, as in Eq(\ref{vev}) and $\pi $ field is massless (Goldstone boson).
Equivalently, we can use a complex field defined by $\phi =\frac{1}{\sqrt{2}}%
\left( \sigma +i\pi \right) $ so that the Lagrangian is of the form 
\begin{equation}
\mathcal{L}=\partial _{\mu }\phi ^{\dagger }\partial ^{\mu }\phi +\mu
^{2}\phi ^{\dagger }\phi -\lambda \left( \phi ^{\dagger }\phi \right) ^{2}
\end{equation}
The symmetry transformation is then $\phi \rightarrow \phi ^{\prime
}=e^{i\alpha }\phi ,$ $\alpha $ is some constant .Now we use the polar
coordinates for the complex field 
\begin{equation}
\phi \left( x\right) =\frac{1}{\sqrt{2}}\left( \rho \left( x\right)
+v\right) \exp \left( \frac{i\theta \left( x\right) }{v}\right)
\end{equation}
to write the Lagrangian in the form, 
\begin{equation}
\mathcal{L}=\frac{1}{2}\left( \partial _{\mu }\rho \right) ^{2}+\frac{\left(
\rho +v\right) ^{2}}{2v^{2}}\left( \partial _{\mu }\theta \right) ^{2}+\frac{%
\mu ^{2}}{2}\left( \rho +v\right) ^{2}-\frac{\lambda }{4}\left( \rho
+v\right) ^{4}
\end{equation}
This clearly shows that $\theta \left( x\right) $ is massless and has only
derivative couplings. This follows from the fact that the $U\left( 1\right) $
(or $SO\left( 2\right) $) symmetry $\phi \rightarrow e^{i\alpha }\phi ,$
corresponds to $\theta \rightarrow \theta +v\alpha ,$ which is
inhomogeneous. So $\theta \left( x\right) $ needs to have a derivative in
order to be invariant under such inhomogeneous transformation. Note that
this Lagrangian, due to the presence of terms like $\left( \partial _{\mu
}\theta \right) ^{2}\rho ^{2}$ is not renormalizable. But this Lagrangian
will be used only as an effective theory to study the low energy
phenomenology while the renomalizability deals with high energy behavior.%
\newline

\subsection{Goldstone Theorem}

From Eq(\ref{ssbv}) we see that the pions $\stackrel{\rightarrow }{\pi }$
are massless. This is a consequence of the Goldstone theorem which states
that spontaneous breaking of a continuous symmetry will give a massless
particle or zero energy excitation. We will first illustrate this by showing
that quadratic terms in $\stackrel{\rightarrow }{\pi }$ are absent in the
tree level as a consequence of spontaneous symmetry breaking of the original
chiral symmetry.. We will use the $O\left( 4\right) $ notation, $\phi
^{i}=\left( \pi _{1},\pi _{2},\pi _{3,}\sigma \right) $The invariance under
the chiral transformation implies that \quad 
\begin{equation}
\delta V=\frac{\partial V}{\partial \phi _{i}}\delta \phi _{i}=\frac{%
\partial V}{\partial \phi _{i}}\varepsilon _{ij}\phi _{j}=0  \label{invV}
\end{equation}
Differentiating Eq(\ref{invV}) with respect to $\phi _{k}$ and then
evaluating this at the minimum, we see that 
\begin{equation}
\quad \frac{\partial ^{2}V}{\partial \phi _{i}\partial \phi _{k}}|_{\min
}\varepsilon _{ij}\langle \phi _{j}>=0,
\end{equation}
For the case $<\phi _{j}>=\delta _{i4}v,$ we see that in the expansion of $V$
around the minimum, $\sigma =v,\pi _{i}=0,$ there are no terms of the form, $%
\pi _{i}\pi _{j},\sigma ^{\prime }\pi _{i},$ with $\sigma ^{\prime }=\sigma
-v.$ Therefore $\pi _{i}^{\prime }s$ are massless in the tree level. We can
extend this to more general case where the effective potential is written as 
$V\left( \phi _{i}\right) .$ This potential is invariant under some symmetry
group $G,$ which transforms $\phi _{i}$ as 
\begin{equation}
\phi _{i}\rightarrow \phi _{i}^{^{\prime }}=\phi _{i}+\alpha
^{a}t_{ij}^{a}\phi _{j}\quad \text{or\quad }\delta \phi _{i}=\alpha
^{a}t_{ij}^{a}\phi _{j}
\end{equation}
where $|\alpha _{i}|\ll 1$ are the parameters for the infinitesimal
transformations and $t^{a}$ matrix for the representation where $\phi _{i}$
belongs. The invariance under these transformation implies that 
\begin{equation}
\frac{\partial V}{\partial \phi _{i}}\alpha ^{a}t_{ij}^{a}\phi _{j}=0
\label{minimum}
\end{equation}
The minimum of the potential is located at $\phi _{i}=v_{i,}$ which
satisfies the equation, 
\begin{equation}
\left( \frac{\partial V}{\partial \phi _{i}}\right) _{\phi =v}=0
\end{equation}
Differentiating Eq (\ref{minimum}) with respect to $\phi _{k}$ and
evaluating this at the minimum, $\phi _{i}=v_{i,}$ we get 
\begin{equation}
\left( \frac{\partial ^{2}V}{\partial \phi _{k}\partial \phi _{i}}\right)
_{\phi =v}\left( t_{ij}^{a}v_{j}\right) =0
\end{equation}
This means that the vector $u_{i}^{a}=t_{ij}^{a}v_{j},$ if non-zero, is an
eigenvector of the mass matrix 
\begin{equation}
m_{ij}^{2}=\left( \frac{\partial ^{2}V}{\partial \phi _{k}\partial \phi _{i}}%
\right) _{\phi =v}
\end{equation}
with zero eigenvalue(massless). Thus the number of massless Goldstone bosons
is just the number of independent vectors of the form, $%
u_{i}^{a}=t_{ij}^{a}v_{j}.$ In other words, if $u_{i}^{a}\neq 0,$ the
combination 
\begin{equation}
\chi ^{a}=\stackunder{ij}{\sum }\phi _{i}t_{ij}^{a}v_{j}
\end{equation}
is the Goldstone boson, up to a normalization constant.

These arguments only show that $\stackrel{\rightarrow }{\pi }$'s are
massless in the tree level. It turns out that this property is true
independent of perturbation theory and can be illustrated in case of $\sigma 
$-model as follows. The axial charge is of the form, 
\begin{equation}
Q_{i}^{5}=i\int d^{3}x\left[ \pi _{i}\partial _{0}\sigma -\sigma \partial
_{0}\pi _{i}+\cdots \right]
\end{equation}
$\prod $Since $\partial _{0}\pi _{i}$ and $\partial _{0}\sigma $ are just
the momenta conjugate to $\pi _{i}$ and $\sigma ,$ we can derive, 
\begin{equation}
\left[ Q_{i}^{5},\pi _{j}\left( 0\right) \right] =\delta _{ij}\sigma \left(
0\right) .
\end{equation}
Between vacuum states, this yields 
\begin{equation}
<0|\left[ Q_{i}^{5},\pi _{j}\left( 0\right) \right] |0>=\delta
_{ij}<0|\sigma \left( 0\right) |0>.  \label{goldstone}
\end{equation}
For the case of SSB, we have 
\begin{equation}
<0|\sigma \left( 0\right) |0>\neq 0
\end{equation}
Note that this condition implies that the axial charges $Q_{i}^{5}$'s do not
annihilate the vacuum, $Q_{i}^{5}|0>\neq 0.$Using $Q_{i}^{5}=\int
A_{i}^{0}\left( x\right) d^{3}x$ we can write the LHS of Eq (\ref{goldstone}%
) as 
\begin{eqnarray}
\langle 0|[Q_{i}^{5},\pi _{j}\left( 0\right) ]|0\rangle &=&i\int
d^{3}x\langle 0|\left[ A_{i}^{0}\left( x\right) ,\pi _{j}\right] |0\rangle =%
\stackunder{n}{i\sum }\delta ^{3}\left( \stackrel{\rightarrow }{p_{n}}%
\right) [\langle 0|A_{i}^{0}\left( 0\right) |n\rangle  \nonumber \\
&&\langle n|\pi _{j}|0\rangle e^{-iE_{n}t}-\langle n|\pi _{j}|0\rangle
\langle 0|A_{i}^{0}\left( x\right) |n\rangle e^{iE_{n}t}]
\end{eqnarray}
This has explicit dependence on $t,$ while the right hand side, $<0|\sigma
|0>,$ is independent of time. The only way these two features can be
consistent is to have a state with the property that 
\begin{equation}
E_{n}\rightarrow 0\quad \text{as\quad }\overrightarrow{p_{n}}\rightarrow 0.
\label{massless}
\end{equation}
This is the content of the Goldstone theorem. For the relativistic system,
the energy and momentum is related by $E_{n}=\sqrt{\stackrel{\rightarrow 2}{%
p_{n}}+m_{n}^{2}}.$Then Eq (\ref{massless}) implies the existence of
massless particle in the system. More specifically, there are physical
states $|\pi _{l}\rangle $ with the property that 
\[
\langle 0|A_{i}^{0}\left( 0\right) |\pi _{l}\rangle \langle \pi _{l}|\pi
_{j}|0\rangle \neq 0 
\]
and are massless from Goldstone theorem. It is convenient to choose the
normalization such that $<\pi _{l}|\pi _{j}|0>=\delta _{ij}$ and write 
\begin{equation}
\langle 0|A_{i}^{\mu }\left( 0\right) |\pi _{l}\left( p\right) \rangle
=if_{\pi }p^{\mu }\delta _{il}\quad \text{with }f_{\pi }\text{ a constnat}
\end{equation}
It is easy to see that 
\begin{equation}
f_{\pi }=\langle 0|\sigma |0\rangle
\end{equation}
For the non-relativistic system Eq(\ref{massless}) simply says that the
dispersion relation $E\left( p\right) $has zero energy excitation.

\subsection{Non-linear $\sigma $-model}

In the $\sigma $-model without the nucleons, we have 3 massless $\pi $'s and
a massive $\sigma $ field. For the energies much smaller than $m_{\sigma },$
the massless Goldstone bosons are the important physical degrees of freedoms
and it is desirable to write down an effective theory with $\pi $'s only. As
we have seen, the theory with SSB has many physical consequences, e.g. low
energy theorem for the Goldstone bosons. The removal of $\sigma $-field
should preserve symmetry so that these results are maintained. Also,
phenomenologically there are no good evidence for the existence of the $%
\sigma $ meson which is the partner of $\pi ^{\prime }s$ in the chiral
symmetry. We now discuss the explicit steps for carrying out this process.
Write the scalar fields as a vector in 4-dimensional space, 
\begin{equation}
\phi _{i}=\left( \phi _{1,}\phi _{2,}\phi _{3,}\phi _{4,}\right) =\left( 
\stackrel{\rightarrow }{\pi },\sigma \right)
\end{equation}
We want to parametrize the $\phi $ fields in such a way that the
non-Goldstone field to be eliminated later is $O\left( 4\right) $ invariant.
One simple parametrization for this purpose is 
\begin{equation}
\phi _{i}=R_{i4}\left( x\right) s\left( x\right) ,\qquad i=1,\cdots 4
\end{equation}
where $R_{ab}$ a $4\times 4$ orthogonal matrix, $RR^{T}=R^{T}R=1,$ which
gives $R_{i4}R_{i4}=1$ and 
\begin{equation}
\phi _{i}\phi _{i}=s^{2}.
\end{equation}
So $s\left( x\right) $ is the magnitude of the vector $\phi _{i}$ and is
clearly $O\left( 4\right) $ invariant. Thus it can be eliminated without
effecting the symmetry. One simple choice for $R_{i4\text{ }}$is, (\cite
{weinberg2}) 
\begin{equation}
R_{a4}\left( x\right) =\frac{2\eta _{a}\left( x\right) }{1+\overrightarrow{%
\eta ^{2}}},\qquad a=1,2,3\qquad R_{44}\left( x\right) =\frac{1-%
\overrightarrow{\eta ^{2}}}{1+\overrightarrow{\eta ^{2}}}  \label{s3para}
\end{equation}
Note that we can invert these relations to get 
\begin{equation}
\stackrel{\rightarrow }{\eta }=\frac{\stackrel{\rightarrow }{\pi }}{\sigma +s%
}  \label{s3para2}
\end{equation}
The Lagrangian is of the form 
\begin{equation}
\mathcal{L}=\frac{1}{2}\left[ \left( \partial _{\mu }s\right) ^{2}+4s^{2}%
\frac{\left( \partial _{\mu }\stackrel{\rightarrow }{\eta }\right) ^{2}}{%
\left( 1+\stackrel{\rightarrow }{\eta }^{2}\right) ^{2}}\right] +\frac{1}{2}%
\mu ^{2}s^{2}-\frac{\lambda }{4}s^{4}
\end{equation}
So $\eta _{i}^{^{\prime }}s$ are the massless Goldstone bosons. To study the
physics of Goldstone bosons at low energies, $E\ll m_{\sigma }$, we can
replace the $s$ field by a constant, $s\left( x\right) =v$ to get 
\begin{equation}
\mathcal{L}=2v^{2}\frac{\left( \partial _{\mu }\stackrel{\rightarrow }{\eta }%
\right) ^{2}}{\left( 1+\stackrel{\rightarrow }{\eta }^{2}\right) ^{2}}
\end{equation}
In order to get the correct nomoralization we rescale, $\stackrel{%
\rightarrow }{\eta }=\frac{\stackrel{\rightarrow }{\pi }^{\prime }}{2v}$ so
that 
\begin{equation}
\mathcal{L}=\frac{1}{2}\frac{\left( \partial _{\mu }\stackrel{\rightarrow }{%
\pi }^{\prime }\right) ^{2}}{\left( 1+\frac{\stackrel{\rightarrow }{\pi }%
^{\prime 2}}{4v^{2}}\right) ^{2}}=\frac{1}{2}\left( \partial _{\mu }%
\stackrel{\rightarrow }{\pi }^{\prime }\right) ^{2}-\frac{1}{4v^{2}}\left( 
\stackrel{\rightarrow }{\pi }^{\prime }\right) ^{2}\left( \partial _{\mu }%
\stackrel{\rightarrow }{\pi }^{\prime }\right) ^{2}+\cdots \cdots
\label{nonlinL}
\end{equation}
Here the interaction terms will always contain derivatives and amplitudes
involving $\stackrel{\rightarrow }{\pi }^{\prime }s$ will vanish in the
limit of zero momenta (low energy theorem). According to the theorem of
field redefinition, this describes the same physics as the usual $\sigma $%
-model Lagrangian in the Goldstone sector. For example, we can check that in
the simple case of $\pi ^{0}\pi ^{0}$ scattering in the tree level the
amplitude from this Lagrangian in Eq (\ref{nonlinL}) is 
\begin{equation}
M=\frac{2i}{v^{2}}\left[ p_{1}\cdot p_{2}-p_{1}\cdot p_{3}-p_{1}\cdot
p_{4}\right] =\frac{i}{v^{2}}\left( s+t+u\right) .
\end{equation}
This is the same result as in Eq(\ref{pipi} ), obtained in the $\sigma $%
-model with $m_{\sigma }\rightarrow \infty .$

The Lagrangian in Eq (\ref{nonlinL}) which contains on the Goldstone boson
fields, is one example of non-linear realization of chiral symmetry, which
will be discussed in detail in the next section. Here we want to mention a
useful geometric interpretation of Lagrangian in Eq (\ref{nonlinL}). When we
eliminate the $O\left( 4\right) $ invariant field by setting $s\left(
x\right) =v$ , $\phi _{i}^{^{\prime }}s$ satisfy the relation, 
\[
\phi _{1}^{2}+\phi _{2}^{2}+\phi _{3}^{2}+\phi _{4}^{2}=v^{2} 
\]
which is just the sphere with radius $v$ in 4-dimensional Euclidean space, $%
S^{3}.$ The variables $\eta _{1},\eta _{2\text{ }},$ and $\eta _{3}$ are
just one particular choice of the coordinates of the space $S^{3}.$ The
transformation of Lagrangian from $\phi $ fields to $\eta $ fields can be
understood in terms of metric tensor in $S^{3}.$ For simplicity, consider
just the kinetic terms in $\mathcal{L}$ , 
\[
\mathcal{L}=\frac{1}{2}\left( \partial _{\mu }\phi ^{i}\right) \left(
\partial _{\mu }\phi ^{j}\right) g_{ij} 
\]
where $g_{ij}=\delta _{ij}$ is the trivial metric in the 4-dimensional
Euclidean space. Then the transformation $\phi ^{i}\rightarrow \phi
^{i}\left( \eta \right) $ gives 
\begin{equation}
\partial _{\mu }\phi ^{i}=\frac{\partial \phi ^{i}}{\partial \eta ^{a}}%
\partial _{\mu }\eta ^{a}
\end{equation}
and 
\begin{equation}
\mathcal{L}=\frac{1}{2}\delta _{ij}\frac{\partial \phi ^{i}}{\partial \eta
^{a}}\frac{\partial \phi ^{j}}{\partial \eta ^{b}}\left( \partial _{\mu
}\eta ^{a}\right) \left( \partial ^{\mu }\eta ^{b}\right) =\frac{1}{2}%
g_{ab}\left( \eta \right) \left( \partial _{\mu }\eta ^{a}\right) \left(
\partial ^{\mu }\eta ^{b}\right)  \label{metricL}
\end{equation}
where 
\begin{equation}
g_{ab}\left( \eta \right) =g_{ij}\frac{\partial \phi ^{i}}{\partial \eta ^{a}%
}\frac{\partial \phi ^{j}}{\partial \eta ^{b}}=\frac{4v^{2}\delta _{ab}}{%
\left( 1+\eta ^{2}\right) ^{2}}  \label{metrics3}
\end{equation}
is the induced metric on $S^{3}.$ Thus in the Lagrangian the coefficient of $%
\left( \partial _{\mu }\eta ^{a}\right) \left( \partial ^{\mu }\eta
^{b}\right) $ is the metric of the space $S^{3},$ which is just the coset
space $O\left( 4\right) /O\left( 3\right) .$ In the general case where the
symmetry breaking is of the form, $G\rightarrow H,$ the non-linear
Lagrangian can be written down with the metric on the manifold $G/H$ as in
Eq ( \ref{metricL}).

It is interesting to see how the transformations of $SU\left( 2\right)
\times SU\left( 2\right) $ are realized on this manifold $S^{3}.$ For the
infinitesimal isospin rotation, we have 
\begin{equation}
\delta \stackrel{\rightarrow }{\pi }=\stackrel{\rightarrow }{\alpha }\times 
\stackrel{\rightarrow }{\pi },\qquad \delta \sigma =0\qquad \alpha :\text{%
group parameters}
\end{equation}
which implies from Eq(\ref{s3para2}) that 
\begin{equation}
\delta \stackrel{\rightarrow }{\eta }=\stackrel{\rightarrow }{\alpha }\times 
\stackrel{\rightarrow }{\eta }
\end{equation}
This is just a rotation on the vector $\stackrel{\rightarrow }{\eta }$ and
the Lagrangian in Eq(\ref{metric L}) with metric given in Eq(\ref{metrics3})
is clearly invariant under such transformation. The axial transformation on $%
\stackrel{\rightarrow }{\pi }$ and $\sigma $ is of the form 
\begin{equation}
\delta \stackrel{\rightarrow }{\pi }=\stackrel{\rightarrow }{\beta }\sigma
,\qquad \delta \sigma =-\stackrel{\rightarrow }{\beta }\cdot \stackrel{%
\rightarrow }{\pi },\qquad \stackrel{\rightarrow }{\beta }:\text{group
parameters}
\end{equation}
which gives 
\begin{equation}
\delta \stackrel{\rightarrow }{\eta }=\frac{\stackrel{\rightarrow }{\beta }}{%
2}\left( 1-\eta ^{2}\right) +\stackrel{\rightarrow }{\eta }\left( \stackrel{%
\rightarrow }{\beta }\cdot \stackrel{\rightarrow }{\eta }\right) .
\end{equation}
This transformation is non-linear and inhomogeneous.. But we can get simple
transformation for the combination, 
\begin{equation}
\delta \left( \frac{\partial _{\mu }\stackrel{\rightarrow }{\eta }}{1+\eta
^{2}}\right) =\left( \stackrel{\rightarrow }{\eta }\times \stackrel{%
\rightarrow }{\beta }\right) \times \left( \frac{\partial _{\mu }\stackrel{%
\rightarrow }{\eta }}{1+\eta ^{2}}\right)
\end{equation}
This looks very much like an isospin rotation except that the parameters for
the rotation now depend on the fields $\eta $ and it is easy now to see that
the Lagrangian in Eq(\ref{metricL}) is invariant under the axial
transformations.\newline
\underline{Remark}: We can transform the metric in Eq(\ref{metrics3}) into
the more familiar Robertson-Walker metric used in cosmology as follows.
First we use the spherical coordinates for $\stackrel{\rightarrow }{\eta }$
to write the metric in the line element as 
\begin{equation}
\left( dl\right) ^{2}=g_{ab}d\eta ^{a}d\eta ^{b}=\frac{4v^{2}}{\left( 1+\eta
^{2}\right) ^{2}}\left[ \left( d\eta \right) ^{2}+\eta ^{2}\left( d\theta
\right) ^{2}+\eta ^{2}\sin ^{2}\theta \left( d\phi \right) ^{2}\right] . 
\nonumber
\end{equation}
Define the new variable $r$ by 
\begin{equation}
r=\frac{2\eta }{1+\eta ^{2}}.
\end{equation}
In terms of new variable the line element is of form,

\begin{equation}
\left( dl\right) ^{2}=\frac{\left( dr\right) ^{2}}{1-r^{2}}+r^{2}\left(
d\theta \right) ^{2}+r^{2}\sin ^{2}\theta \left( d\phi \right) ^{2}
\end{equation}
which is just the usual Robertson-Walker metric for the case of positive
curvature.

\section{Non-linear Realization}

As we have seen in the last section, it is useful to write down a Lagrangian
with only Goldstone bosons as an effective theory to describe physics at low
energies. In this section we will discuss the general description of this
procedure, which is usually called the non-linear realization. The usual
discussion of this subject is generally rather formal and abstract. The
discussion here will emphasize the intuitive understanding rather than the
mathematical rigor.

In order to make the discussion here somewhat self contained, we will first
discuss some simple results from group theory (\cite{group}) which are
useful for the understanding of non-linear realization. Then we discuss the
general features of non-linear realization.

\subsection{Useful results from group theory}

Here we will recall the rearrangement theorem which is central to most of
the group theoretical result and then discuss the concept of coset space
which forms the basis of the non-linear realization.

$\left( i\right) $\textbf{Rearrangement Theorem}\newline
Let $G=\{g_{1},\cdots g_{n}\}$ be a finite group. If we multiply the whole
group by an arbitrary group element $g_{i},$i.e. $\{g_{i}g_{1},\cdots
g_{i}g_{n}\},$ the resulting set is just the group $G$ itself.\newline

$\left( ii\right) $\textbf{Coset space}\newline
Coset space decomposes a group into non-overlapping sets with respect to a
subgroup. Let $H=\{h_{1,\cdots }h_{l}\}$ be a non-trivial subgroup of $G$.
For any element $g_{i}$ in $G$ but not in $H$, the left coset $g_{i}H$, or
coset for short, is just $\{g_{i}h_{1,\cdots }g_{i}h_{l}\}.$ The coset $%
g_{i}H$ will not have any element in common with the subgroup $H,$ and any
two such cosets are either identical or have no elements in common. This can
be seen as follows. Consider cosets $g_{1}H,$ and $g_{2}H.$ Suppose that
there is one element in common, 
\[
g_{1}h_{i}=g_{2}h_{j}\qquad \text{for some }i,j 
\]
Then we can write 
\[
g_{1}^{-1}g_{2}=h_{i}h_{j}^{-1} 
\]
which means $g_{1}^{-1}g_{2}$ is one of the element of subgroup $H.$ Then by
the rearrangement theorem applied to the subgroup $H,$ we get 
\[
g_{1}^{-1}g_{2}H=H,\qquad \Rightarrow \qquad g_{1}H=g_{2}H 
\]
i.e. these two cosets have the same group elements. The group $G$ is now
decomposed into these non-overlapping cosets and the collection of all the
distinct cosets $g_{1}H,\cdots g_{k}H,$ together with $H,$ will contain all
the group elements of $G.$ This is denoted by $G/H.$ This can be generalized
to Lie group where rearrangement theorem is valid. \newline
\underline{Example of coset space}: Consider points on 2-dimensional plane
which form a group under the addition, 
\[
\left( x_{1},y_{1}\right) \cdot \left( x_{2},y_{2}\right) =\left(
x_{1}+x_{2},y_{1}+y_{2}\right) 
\]
Clearly, points on the $y$-axis, $\left( 0,y\right) ,$ form a subgroup,
denoted by $H.$ Then the vertical line of the form, $\left( a,y\right) $with
a fixed is a coset with respect to the subgroup $H.$ It is clear that the
whole 2-dimensional plane can be decomposed into collection of such vertical
lines. We can label these cosets, vertical lines, by choosing one element
from each coset. Clearly there are many ways to choose such representatives.
One convenient parametrization is to choose those points on the $x$-axis, so
that the cosets are of the form $x_{i}H.$ Each group element $\left(
x,y\right) $ can be written as the product, 
\[
\left( x,y\right) =\left( x,0\right) \cdot \left( 0,y\right) 
\]
where $\left( 0,y\right) \in H.$ Under the action of an arbitrary group
element $g=\left( a,b\right) $ this will give 
\begin{eqnarray*}
g\left( x,y\right) &=&\left( a,b\right) \cdot \left( x,y\right) =\left(
a,b\right) \cdot \left( x,0\right) \cdot \left( 0,y\right) =\left(
a+x,0\right) \cdot \left( 0,b\right) \cdot \left( 0,y\right) \\
&=&\left( a+x,0\right) \cdot \left( 0,b+y\right)
\end{eqnarray*}
(the computation here is organized in such a way that it parallel to the
more complicate case in the non-linear realization.) Thus the group element $%
g=\left( a,b\right) $ will move the points in the coset $xH$ to points in
the coset $\left( a+x\right) H.$ In terms of coset parameters, we have 
\[
g:x\rightarrow x+a. 
\]

\subsection{Non-linear Realization of Symmetries}

We will first discuss the general machinery of non-linear realization (\cite
{coleman1}\cite{coleman2}) and then take up the special case of chiral
symmetry where the parity symmetry will make the realization much simpler..

\subsubsection{General case}

Suppose the symmetry group $G$ is spontaneously broken to a subgroup $H.$
where both $G$ and $H$ are Lie groups. Choose the generators of $G$ to be of
the form $\left\{ V_{1},,\ldots V_{l},\text{ }A_{1,}\text{ }\ldots
A_{k}\right\} $ such that $\left\{ V_{1},V_{2},\ldots V_{l},\right\} $ are
the generators of the subgroup $H.$ As usual, the group elements in $H$ can
be written in the form, 
\[
\exp \left( i\overrightarrow{\alpha }\cdot \overrightarrow{V}\right) 
\]
where $\alpha _{1},\alpha _{2},\ldots ,\alpha _{l},$ are the group
parameters for $H$ and are taken to be real. From the coset decomposition,
we can write an arbitrary group element in $G$ as 
\begin{equation}
g=e^{i\overrightarrow{\xi }\cdot \overrightarrow{A}}e^{i\overrightarrow{%
\alpha }\cdot \overrightarrow{V}}  \label{g}
\end{equation}
Here $\xi _{1},\cdots \xi _{l},$are the parameters which label the coset $%
e^{i\overrightarrow{\xi }\cdot \overrightarrow{A}}H.$ Note that since the
vacuum is invariant under $H,$ we have 
\begin{equation}
g|0>=e^{i\overrightarrow{\xi }\cdot \overrightarrow{A}}e^{i\overrightarrow{%
\alpha }\cdot \overrightarrow{V}}|0>=e^{i\overrightarrow{\xi }\cdot 
\overrightarrow{A}}|0>
\end{equation}
i.e. $\overrightarrow{\xi }$ also labels the different vacua, which are
degenerate. Recall that the different vacua form the manifold $G/H.$ Thus
the coset parameters are also the parameters for the manifold $G/H.$ Under
the action of an arbitrary group element $g_{1}\in G,$ we have the
combination 
\begin{equation}
g_{1}g=g_{1}e^{i\overrightarrow{\xi }\cdot \overrightarrow{A}}e^{i%
\overrightarrow{\alpha }\cdot \overrightarrow{V}}.
\end{equation}
Since $g_{1}e^{i\overrightarrow{\xi }\cdot \overrightarrow{A}}$ is also a
group element in $G$, we can write a coset decomposition, 
\begin{equation}
g_{1}e^{i\overrightarrow{\xi }\cdot \overrightarrow{A}}=e^{i\overrightarrow{%
\xi }^{\prime }\cdot \overrightarrow{A}}e^{i\overrightarrow{\alpha }^{\prime
}\cdot \overrightarrow{V}}
\end{equation}
and then 
\begin{equation}
g_{1}\left( e^{i\overrightarrow{\xi }\cdot \overrightarrow{A}}e^{i%
\overrightarrow{\alpha }\cdot \overrightarrow{V}}\right) =e^{i%
\overrightarrow{\xi }^{\prime }\cdot \overrightarrow{A}}e^{i\overrightarrow{%
\alpha }^{\prime }\cdot \overrightarrow{V}}e^{i\overrightarrow{\alpha }\cdot 
\overrightarrow{V}}=e^{i\overrightarrow{\xi }^{\prime }\cdot \overrightarrow{%
A}}e^{i\overrightarrow{\alpha }^{\prime \prime }\cdot \overrightarrow{V}}
\end{equation}
where we have used the group property of $H$ to write 
\begin{equation}
e^{i\overrightarrow{\alpha }^{\prime }\cdot \overrightarrow{V}}e^{i%
\overrightarrow{\alpha }\cdot \overrightarrow{V}}=e^{i\overrightarrow{\alpha 
}^{\prime \prime }\cdot \overrightarrow{V}}
\end{equation}
Note that the new coset parameters $\overrightarrow{\xi ^{\prime }}$ and
parameters $\stackrel{\rightarrow }{\alpha }^{\prime }$for the subgroup $H,$%
all depend on the original coset parameters $\overrightarrow{\xi },$%
\begin{equation}
\overrightarrow{\xi ^{\prime }}=\overrightarrow{\xi ^{\prime }}\left( 
\overrightarrow{\xi },g_{1}\right) ,\qquad \overrightarrow{\alpha ^{\prime }}%
=\overrightarrow{\alpha ^{\prime }}\left( \overrightarrow{\xi },g_{1}\right)
\label{coset}
\end{equation}
In this way, the group element $g_{1}$ transform the coset parameters from $%
\overrightarrow{\xi }\rightarrow \overrightarrow{\xi ^{\prime }}$ in the
coset space $G/H.$ As we will see later, these coset parameters will be
identified with the Goldstone bosons. These transformations on $%
\overrightarrow{\xi },$ and $\overrightarrow{\alpha }$ induced by the group
elements will have the same group properties as the group elements and are
called the \textit{non-linear realization }of the group. This is in contrast
to the usual representation of the group where group elements are
represented by matrices. In the transformation in Eq(\ref{coset}) $%
\overrightarrow{\xi ^{\prime }}$ is generally not a linear function of $%
\overrightarrow{\xi }.$ But for the special case where $g=h$ is a group
element from the unbroken subgroup $H,$ we get, from Eq(\ref{g}), 
\begin{equation}
hg=he^{i\overrightarrow{\xi }\cdot \overrightarrow{A}}e^{i\overrightarrow{%
\alpha }\cdot \overrightarrow{V}}=\left( he^{i\overrightarrow{\xi }\cdot 
\overrightarrow{A}}h^{-1}\right) \left( he^{i\overrightarrow{\alpha }\cdot 
\overrightarrow{V}}\right) =\left( he^{i\overrightarrow{\xi }\cdot 
\overrightarrow{A}}h^{-1}\right) e^{i\overrightarrow{\alpha }\cdot 
\overrightarrow{V^{\prime }}}
\end{equation}
where 
\begin{equation}
he^{i\overrightarrow{\alpha }\cdot \overrightarrow{V}}=e^{i\overrightarrow{%
\alpha }\cdot \overrightarrow{V^{\prime }}}
\end{equation}
In general, the broken generators $\stackrel{\rightarrow }{A}$ transform as
some representation $D$ with respect to the subgroup $H,$%
\begin{equation}
hA_{i}h^{-1}=A_{j}D_{ji}\left( h\right)
\end{equation}
For example, in the case of $G=SU\left( 2\right) \times SU\left( 2\right) $
model, the broken generators, $A_{1},A_{2},A_{3}$ transform as triplet under
the unbroken subgroup $H=SU\left( 2\right) .$ We can then write 
\begin{equation}
he^{i\overrightarrow{\xi }\cdot \overrightarrow{A}}h^{-1}=\exp \left( i\xi
_{i}hA_{i}h^{-1}\right) =\exp \left( i\xi _{i}A_{j}D_{ji}\left( h\right)
\right) =\exp \left( i\xi _{j}^{^{\prime }}A_{j}\right)
\end{equation}
where 
\begin{equation}
\xi _{j}^{^{\prime }}=D_{ji}\left( h\right) \xi _{i}  \label{linearrep}
\end{equation}
This means that $\xi _{i}^{\prime }s$ transform linearly under the subgroup $%
H.$ Also it is easy to see that the parameters $\stackrel{\rightarrow }{%
\alpha }$ are independent of the coset parameters $\xi _{i}$. But if the
group element is of the form $e^{i\overrightarrow{\xi ^{\prime }}\cdot 
\overrightarrow{A}}$ the transformation law for the coset parameters $\xi $
is non-linear and quite complicate.

\subsubsection{Chiral symmetry}

For the case of chiral symmetry, there is significant simplification due to
parity operation. We will illustrate this in the simple case of $SU\left(
2\right) _{L}\times SU\left( 2\right) _{R}$ symmetry. The parity operation
is of the form, 
\[
P:\quad \stackrel{\rightarrow }{V}\rightarrow \stackrel{\rightarrow }{V}%
,\quad \stackrel{\rightarrow }{A}\rightarrow -\stackrel{\rightarrow }{A} 
\]
Consider the case where group element $g$ consists of left-handed
transformation, 
\begin{equation}
g=\exp \left( i\stackrel{\rightarrow }{\theta }\cdot \left( \stackrel{%
\rightarrow }{V}-\stackrel{\rightarrow }{A}\right) \right) \equiv L
\end{equation}
and write the transformation of coset parameters as 
\begin{equation}
ge^{i\overrightarrow{\xi }\cdot \overrightarrow{A}}=Le^{i\overrightarrow{\xi 
}\cdot \overrightarrow{A}}=e^{i\overrightarrow{\xi ^{\prime }}\cdot 
\overrightarrow{A}}e^{i\overrightarrow{\alpha ^{\prime }}\cdot 
\overrightarrow{V}}  \label{l}
\end{equation}
Under the parity transformation, the left-handed transformation is changed
into right-handed one, 
\begin{equation}
P\left( L\right) =P\left( e^{i\stackrel{\rightarrow }{\theta }\cdot \left( 
\stackrel{\rightarrow }{V}-\stackrel{\rightarrow }{A}\right) }\right) =e^{i%
\stackrel{\rightarrow }{\theta }\cdot \left( \stackrel{\rightarrow }{V}+%
\stackrel{\rightarrow }{A}\right) }\equiv R
\end{equation}
Then applying the parity transformation to Eq(\ref{l}), we get 
\begin{equation}
R\,e^{-i\overrightarrow{\xi }\cdot \overrightarrow{A}}=e^{-i\overrightarrow{%
\xi ^{\prime }}\cdot \overrightarrow{A}}e^{i\overrightarrow{\alpha ^{\prime }%
}\cdot \overrightarrow{V}}  \label{r}
\end{equation}
Using the notation 
\begin{equation}
\Sigma \equiv e^{i\overrightarrow{\xi }\cdot \overrightarrow{A}},\qquad
h=e^{i\overrightarrow{\alpha ^{\prime }}\cdot \overrightarrow{V}}
\end{equation}
we can combine the Eqs(\ref{l},\ref{r}) into 
\begin{equation}
\Sigma ^{\prime }=L\Sigma h^{\dagger }=h\Sigma R^{\dagger }  \label{linearr}
\end{equation}
Note that as we have mentioned before, the parameters $\overrightarrow{%
\alpha ^{\prime }}$ depend on the coset parameters $\overrightarrow{\xi },$
the transformation law for $\Sigma $ here is non-linear because the factor $%
h $ depends on $\stackrel{\rightarrow }{\alpha }^{\prime }$. However, the
combination $U=\Sigma ^{2}$ will have a simple transformation law, 
\begin{equation}
U^{\prime }=LUR^{\dagger }.  \label{linear}
\end{equation}
Since $L,R$ are independent of the coset parameters $\overrightarrow{\xi },$%
this transformation law is linear and will be useful for constructing
Lagrangian.

We are interested in the cases where spontaneous symmetry breaking is
generated by the scalar fields. Some of these scalar fields become the
massless Goldstone bosons, like the pions in the $\sigma $-model and others
remain massive, like $\sigma $-field. Consider the scalar fields in the $%
\sigma $-model, where we will use the notation, 
\begin{equation}
\stackrel{\rightarrow }{\phi }=\left( \pi _{1},\pi _{2},\pi _{3},\sigma
\right) =\left( \phi _{1,}\phi _{2,}\phi _{3,}\phi _{4}\right)
\end{equation}
The vacuum expectation value which gives the classical ground state is of
the form, 
\begin{equation}
<\phi >_{0}=\left( 
\begin{array}{c}
0 \\ 
0 \\ 
0 \\ 
v
\end{array}
\right)
\end{equation}
Suppose we make a very general assumption that from a given point in $\phi $%
-space we can reach any other point by some group transformation.(Space is
transitive). Then the general field configuration can be written as 
\begin{equation}
\phi \left( x\right) =\left( 
\begin{array}{c}
\phi _{1}\left( x\right) \\ 
\phi _{2}\left( x\right) \\ 
\phi _{3}\left( x\right) \\ 
\phi _{4}\left( x\right)
\end{array}
\right) =g\left( 
\begin{array}{c}
0 \\ 
0 \\ 
0 \\ 
\sigma
\end{array}
\right) \qquad \text{for some }g\in G
\end{equation}
\underline{Remark}:Strictly speaking we should use the representation
matrices $D\left( g\right) $ of the group element $g$ rather than $g$
itself. However for simplicity of notation, $g$ here is a shorthand for $%
D\left( g\right) .$ From the coset decomposition in Eq(\ref{g}),we can write
\qquad 
\begin{equation}
g=e^{i\overrightarrow{\pi }\cdot \overrightarrow{A}}e^{i\overrightarrow{%
\alpha }\cdot \overrightarrow{V}}
\end{equation}
Here we have chosen coset parameters to be the pion fields. Then we can
write the scalar fields as 
\begin{equation}
\phi \left( x\right) =g\left( 
\begin{array}{c}
0 \\ 
0 \\ 
0 \\ 
\sigma
\end{array}
\right) =e^{i\overrightarrow{\pi }\cdot \overrightarrow{A}}e^{i%
\overrightarrow{\alpha }\cdot \overrightarrow{V}}\left( 
\begin{array}{c}
0 \\ 
0 \\ 
0 \\ 
\sigma
\end{array}
\right) =e^{i\overrightarrow{\pi }\cdot \overrightarrow{A}}\left( 
\begin{array}{c}
0 \\ 
0 \\ 
0 \\ 
\sigma
\end{array}
\right)
\end{equation}
where we have used the fact that the vector $\left( 0,0,0,\sigma \right) $
is proportional to the vacuum configuration $\left( 0,0,0,v\right) $ and is
invariant under the subgroup $H=\{e^{i\overrightarrow{\alpha }\cdot 
\overrightarrow{V}}\}.$ Since the Goldstone bosons are identified as the
coset parameters, they transform the same way as $\stackrel{\rightarrow }{%
\xi }$ in Eq(\ref{coset}), i.e. 
\begin{equation}
g_{1}e^{i\overrightarrow{\pi }\cdot \overrightarrow{A}}=e^{i\overrightarrow{%
\pi ^{\prime }}\cdot \overrightarrow{A}}e^{i\overrightarrow{\alpha ^{\prime }%
}\cdot \overrightarrow{V}}
\end{equation}
where 
\begin{equation}
\stackrel{\rightarrow }{\pi ^{\prime }}=\stackrel{\rightarrow }{\pi ^{\prime
}}\left( \pi ,g_{1}\right) ,\qquad \stackrel{\rightarrow }{\alpha ^{\prime }}%
=\stackrel{\rightarrow }{\alpha ^{\prime }}\left( \pi ,g_{1}\right)
\end{equation}
For the case $g=h\in H,$we have from Eq(\ref{linearrep}), 
\begin{equation}
\pi _{j}^{^{\prime }}=D_{ji}\left( h\right) \pi _{i}
\end{equation}
\newline
\underline{Remark}: The scalar fields here have property that after
separating out the Goldstone bosons, the remainders are proportional to the
vacuum expectation value and is then invariant under the subgroup $H.$ This
is true only for scalar fields in the vector representation in $O\left(
n\right) $ or $SU\left( n\right) $ groups and is not true for scalars in
more general representation. In more general cases, we can separate out the
Goldstone bosons by writing $\phi \left( x\right) $ as 
\begin{equation}
\phi \left( x\right) =e^{i\overrightarrow{\pi }\cdot \overrightarrow{A}}\chi
\left( x\right)
\end{equation}
where $\chi \left( x\right) $ contains all the massive fields. Under the
action of group element $g_{1}$, we have 
\begin{equation}
g_{1}\phi =g_{1}e^{i\overrightarrow{\pi }\cdot \overrightarrow{A}}\chi
\left( x\right) =e^{i\overrightarrow{\pi ^{\prime }}\cdot \overrightarrow{A}%
}e^{i\overrightarrow{\alpha ^{\prime }}\cdot \overrightarrow{V}}\chi \left(
x\right) =e^{i\overrightarrow{\pi ^{\prime }}\cdot \overrightarrow{A}}\chi
^{\prime }\left( x\right)
\end{equation}
where 
\begin{equation}
\chi ^{\prime }\left( x\right) =e^{i\overrightarrow{\alpha ^{\prime }}\cdot 
\overrightarrow{V}}\chi \left( x\right)
\end{equation}
To see how the Goldstone bosons transform under the axial transformation, we
set $L=R^{\dagger }=e^{i\stackrel{\rightarrow }{\theta }\cdot 
\overrightarrow{A}},$in Eq(\ref{linearr}) to get 
\begin{equation}
e^{i2\overrightarrow{\pi ^{\prime }}\cdot \overrightarrow{A}}=e^{i%
\overrightarrow{\theta }\cdot \overrightarrow{A}}e^{i2\overrightarrow{\pi }%
\cdot \overrightarrow{A}}e^{i\overrightarrow{\theta }\cdot \overrightarrow{A}%
}
\end{equation}
To understand this equation better, we first take the infinitesimal
transformation, $|\overrightarrow{\theta }|\ll 1,$%
\begin{equation}
e^{i2\overrightarrow{\pi ^{\prime }}\cdot \overrightarrow{A}}=e^{i2%
\overrightarrow{\pi }\cdot \overrightarrow{A}}+\left( i\stackrel{\rightarrow 
}{\theta }\cdot \stackrel{\rightarrow }{A}\right) e^{i2\overrightarrow{\pi }%
\cdot \overrightarrow{A}}+e^{i2\overrightarrow{\pi }\cdot \overrightarrow{A}%
}\left( i\stackrel{\rightarrow }{\theta }\cdot \stackrel{\rightarrow }{A}%
\right) +\cdots
\end{equation}
and then expand this in powers of $\stackrel{\rightarrow }{\pi }^{\prime }$%
s, 
\[
1+2i\stackrel{\rightarrow }{\pi }^{\prime }\cdot \stackrel{\rightarrow }{A}%
+\cdots =1+2i\stackrel{\rightarrow }{\pi }\cdot \stackrel{\rightarrow }{A}+2i%
\stackrel{\rightarrow }{\theta }\cdot \stackrel{\rightarrow }{A}+\cdots 
\]
Comparing both sides we see that 
\begin{equation}
\stackrel{\rightarrow }{\pi }^{\prime }=\stackrel{\rightarrow }{\pi }+%
\stackrel{\rightarrow }{\theta }+\cdots
\end{equation}
Thus there is a inhomogeneous term in the transformation law for the
Goldstone bosons $\stackrel{\rightarrow }{\pi }$ and this is why Goldstone
bosons have derivative coupling.

Since the transformation law for $U=e^{i2\overrightarrow{\pi }\cdot 
\overrightarrow{A}}$ is linear and simple, it is easier to construct the
chirally invariant interaction in terms of $U$ rather than $\Sigma .$ It is
easy to see that the only invariants without derivatives will involve trace
of some powers of $UU^{\dagger },$which is just an identity matrix. (This
also implies that the Goldstone boson coupling will involve derivatives).
Thus the interaction with lowest numbers of derivative is of the form 
\begin{equation}
\mathcal{L}=tr\left( \partial _{\mu }U\partial ^{\mu }U\right)
\end{equation}

\paragraph{\textbf{Covariant derivative}:}

The parity symmetry in the $\sigma $-model is responsible for getting the
simple combination $U\left( x\right) $ which transforms linearly. For the
more general case where there is no such simplification, to construct
invariant terms involving derivatives is quite complicate because the
non-linear transformation law will involve Golstone boson fields which are
space-time dependent. This means that we need to construct the covariant
derivatives. Futhermore, to exhibit the low energy explicitly we need to
couple Golstone field with derivative to other matter fields. We will now
discuss briefly in the simple case of $\sigma $-model. Write the scalar
fields $\phi \left( x\right) $ in the form, 
\begin{equation}
\phi \left( x\right) =e^{i\overrightarrow{\pi }\left( x\right) \cdot 
\overrightarrow{A}}\left( 
\begin{array}{c}
0 \\ 
0 \\ 
0 \\ 
\sigma \left( x\right)
\end{array}
\right) =\Sigma \left( x\right) \chi \left( x\right)
\end{equation}
with $\Sigma \left( x\right) =$ $e^{i\overrightarrow{\pi }\left( x\right)
\cdot \overrightarrow{A}}.$As before under the action of group element $g,$
we have 
\begin{equation}
g\Sigma \left( x\right) =\Sigma ^{\prime }\left( x\right) h\left( x\right)
\qquad \text{with\qquad }h\left( x\right) =e^{i\stackrel{\rightarrow }{%
\alpha }\cdot \stackrel{\rightarrow }{V}}  \label{cov1}
\end{equation}
Since $\phi $ transforms linearly, we have 
\[
\phi ^{\prime }=g\phi 
\]
which implies 
\begin{equation}
\Sigma ^{\prime }\chi ^{\prime }=g\Sigma \chi =\Sigma ^{\prime }h\chi \qquad 
\text{or \qquad }\chi ^{\prime }=h\chi
\end{equation}
This means that $\chi $ transforms non-linearly because $\stackrel{%
\rightarrow }{\alpha }$ in $h$ depend on $\pi $ fields. Making use of the
simplification for the case of chiral symmetry we have, from transformation
law,(\ref{linearr}) 
\begin{equation}
\partial _{\mu }\Sigma ^{\prime }=L(\partial _{\mu }\Sigma h^{-1}+\Sigma
\partial _{\mu }h^{-1})=(\partial _{\mu }\Sigma h+\Sigma \partial _{\mu
}h)R^{\dagger }
\end{equation}
and 
\begin{equation}
\Sigma ^{\prime -1}\partial _{\mu }\Sigma ^{\prime }=h\left( \Sigma
^{-1}\partial _{\mu }\Sigma \right) h^{-1}+h\partial _{\mu }h^{-1}
\end{equation}
\begin{equation}
\partial _{\mu }\Sigma ^{\prime }\Sigma ^{\prime -1}=h\left( \partial _{\mu
}\Sigma \Sigma ^{-1}\right) h^{-1}-h\partial _{\mu }h^{-1}
\end{equation}
If we define 
\begin{equation}
v_{\mu }=\frac{1}{2}\left[ \Sigma ^{-1}\partial _{\mu }\Sigma -\partial
_{\mu }\Sigma \Sigma ^{-1}\right]  \label{covarv}
\end{equation}
\begin{equation}
a_{\mu }=\frac{1}{2}\left[ \Sigma ^{-1}\partial _{\mu }\Sigma +\partial
_{\mu }\Sigma \Sigma ^{-1}\right]  \label{covara}
\end{equation}
we get 
\[
v_{\mu }^{\prime }=hv_{\mu }h^{-1}+h\partial _{\mu }h^{-1} 
\]
\[
a_{\mu }=ha_{\mu }h^{-1} 
\]
This means that $v_{\mu }$ transforms like ''gauge field'', while $a_{\mu }$
transforms as global adjoint field. Therefore $v_{\mu }$ can be used to
construct the covariant derivative and $a_{\mu }$ is like a global axial
vector field.

\paragraph{Nucleon Field}

The nucleon field in the linear $\sigma $-model has the transformation
properties, 
\[
N_{L}\rightarrow N_{L}^{^{\prime }}=LN_{L},\qquad \text{and \qquad }%
N_{R}\rightarrow N_{R}^{^{\prime }}=RN_{R} 
\]
Thus to couple nucleon fields to the Goldstone bosons, we could write down
the following $SU\left( 2\right) \times SU\left( 2\right) $ invariant
coupling, 
\[
\mathcal{L}_{int}=g\left( \overline{N}_{L}\Sigma N_{R}+\overline{N}%
_{R}\Sigma ^{\dagger }N_{L}\right) . 
\]
However, this is not of the form of derivative coupling which exhibits the
low energy theorem explicitly. For this purpose and general non-linear
realization, we define a new nucleon field by, 
\[
N=e^{i\stackrel{\rightarrow }{\pi }\cdot \stackrel{\rightarrow }{A}}%
\widetilde{N} 
\]
Under the action of the group element $g,$we get 
\[
gN=ge^{i\stackrel{\rightarrow }{\pi }\cdot \stackrel{\rightarrow }{A}}%
\widetilde{N}=e^{i\stackrel{\rightarrow }{\pi }^{\prime }\cdot \stackrel{%
\rightarrow }{A}}e^{i\stackrel{\rightarrow }{\alpha }\cdot \stackrel{%
\rightarrow }{V}}\widetilde{N}=e^{i\stackrel{\rightarrow }{\pi }^{\prime
}\cdot \stackrel{\rightarrow }{A}}\widetilde{N}^{\prime } 
\]
where 
\[
\widetilde{N}^{\prime }=e^{i\stackrel{\rightarrow }{\alpha }\cdot \stackrel{%
\rightarrow }{V}}\widetilde{N}=h\widetilde{N} 
\]
Thus $\widetilde{N}$ transforms according to the representation of the
subgroup $H$ but with the group parameters depend on $\pi $ fields, $%
\stackrel{\rightarrow }{\alpha }\left( \stackrel{\rightarrow }{\pi }%
,g\right) .$ Then from the transformation properties given in Eqs(\ref
{covarv,covara}), we can write down the derivative coupling as 
\[
\mathcal{L}_{N}=\overline{\widetilde{N}}\gamma ^{\mu }\left( i\partial _{\mu
}-v_{\mu }\right) \widetilde{N}+g\overline{\widetilde{N}}\gamma ^{\mu
}a_{\mu }\widetilde{N} 
\]
which will yield the low energy theorem explicitly.

\section{Examples in the Non-relativistic System}

In the frame work of relativistic field theory, e.g. in $SU\left( 2\right)
\times SU\left( 2\right) $ $\sigma $-model, spontaneous symmetry breaking
seems to be put in by hand, i.e. setting the quadratic terms to have
negative sign in the scalar potential in order to develope vacuum
expectation value. This is rather ad hoc and no physical reason is given for
why this is the case. We will now discuss some simple non-relativistic
examples of spontaneous symmetry breaking in order to shed some light on
this (\cite{negele}).

\subsection{Infinite range Ising model}

Consider a system of $N$ spins on an one dimensional lattice with
Hamiltonian, 
\begin{equation}
H=-\frac{J}{N}\stackrel{N}{\stackunder{i<j}{\sum }}s_{i}s_{j}-B\stackrel{N}{%
\stackunder{i}{\sum }}s_{i}
\end{equation}
where $s_{i}=\pm 1,J$ is the coupling constant for spin-spin interaction and 
$B$ is the external magnetic field. In this Hamiltonian, for calculational
simplicity we allow every spin to interact with every other spin, while more
realistic situation will be the short range nearest neighbor interaction.
But the interest here is to see how the spontaneous symmetry breaking come
about and we will ignore this. The partition function is given by 
\begin{equation}
Z=Tr\left( e^{-\beta H}\right) =\stackunder{s_{i}=\pm 1}{\sum }\exp \left( 
\frac{\beta J}{2N}\left( \stackunder{i}{\sum }s_{i}\right) ^{2}+\beta B%
\stackunder{i}{\sum }s_{i}\right)
\end{equation}
Using the identity for the Gaussian integral, 
\[
\int_{-\infty }^{+\infty }dxe^{-ax^{2}+bx}=\sqrt{\frac{\pi }{a}}e^{-\frac{%
b^{2}}{4a}} 
\]
we can write the partition function as 
\begin{equation}
Z=\stackunder{s_{i}=\pm 1}{\sum }\sqrt{\frac{N\beta J}{2\pi }}\int_{-\infty
}^{+\infty }dx\exp \left[ -\frac{N\beta J}{2}x^{2}+\beta \left( Jx+B\right)
\left( \stackunder{i}{\sum }s_{i}\right) \right]
\end{equation}
Now we can sum over each $s_{i}$ independently, 
\[
\stackunder{s_{i}=\pm 1}{\sum }\exp \left( \beta \left( Jx+B\right) \left( 
\stackunder{i}{\sum }s_{i}\right) \right) =\exp \{N\log \left[ 2\cosh \beta
\left( Jx+B\right) \right] \} 
\]
and 
\begin{equation}
Z=\sqrt{\frac{N\beta J}{2\pi }}\int_{-\infty }^{+\infty }dx\exp \left( -%
\frac{N\beta J}{2}x^{2}+N\log \left[ 2\cosh \beta \left( Jx+B\right) \right]
\right)
\end{equation}
From the partition function $Z$, we can compute the average spin, 
\begin{equation}
S=\frac{1}{N}<\stackunder{i}{\sum }s_{i}>=-\frac{1}{\beta N}\frac{\partial }{%
\partial B}\ln Z.
\end{equation}
If $S\neq 0$ in the limit the external field vanishes, $B\rightarrow 0$,
then we have spontaneous symmetry breaking. Since we are interested in the
case where $N$ is very large, we can use saddle point method to compute $Z.$
Write 
\begin{equation}
Z=\sqrt{\frac{N\beta J}{2\pi }}\int_{-\infty }^{+\infty }dx\exp \{-N\beta
f\left( x\right) \}
\end{equation}
where 
\begin{equation}
f\left( x\right) =\frac{Jx^{2}}{2}-\frac{1}{\beta }\log \left[ 2\cosh \beta
\left( Jx+B\right) \right]
\end{equation}
The minimum of $f\left( x\right) $ is given by 
\begin{equation}
f^{\prime }\left( x\right) =0,\qquad \Rightarrow \qquad x=\tanh \beta \left(
Jx+B\right)  \label{minm}
\end{equation}
Let $x_{i},i=0,1,2,\cdots $ be the solutions of this transcendental
equation, then 
\begin{equation}
Z=\sqrt{\frac{N\beta J}{2\pi }}\stackunder{i}{\sum }\exp \{-N\beta f\left(
x_{i}\right) \}\sqrt{\frac{2\pi }{N\beta f^{\prime \prime }\left(
x_{i}\right) }}  \label{partition}
\end{equation}
Suppose $x_{0}$ is the smallest of these solutions, it will dominate the
partition function for large $N.$ Then the average magnetization is then 
\begin{equation}
S=-\frac{1}{\beta N}\frac{\partial }{\partial B}\ln Z=\tanh \beta \left(
Jx_{0}+B\right) =x_{0}
\end{equation}
where we have used the equation satisfied by $x_{0}.$Thus the minimum of $%
f\left( x\right) $ will correspond to the average magnetization.

To study spontaneous symmetry breaking, we set $B=0,$ in Eq(\ref{minm}), and
get 
\begin{equation}
x=\tanh \beta Jx
\end{equation}
It turns out that this equation has only the trivial solution, $x=0$ if $%
\beta J<1$ and non-trivial solution exists only for $\beta J>1.$ To
understand this feature, we expand $f\left( x\right) $ in powers of $x$ for
the case $B=0,$%
\begin{eqnarray}
f\left( x\right) &=&\frac{Jx^{2}}{2}-\frac{1}{\beta }\log \left[ 1+\frac{1}{2%
}\left( \beta Jx\right) ^{2}+\frac{1}{4!}\left( \beta Jx\right) ^{4}+\cdots
\right] \\
&=&\frac{J}{2}\left( 1-\beta J\right) x^{2}+\frac{1}{12}\beta
^{3}J^{4}x^{4}+\cdots  \nonumber
\end{eqnarray}
Thus $\beta J>1$ corresponds to negative quadratic term, which is the
familiar situation in the scalar potential in the $\sigma $-model and the
like. In terms of temperature this condition, we have 
\begin{equation}
J>kT.  \label{critical}
\end{equation}
Here $J$ is the coupling which wants to align the spins in the same
direction while the effect of temperature is to randomize spins. Thus the
condition in Eq(\ref{critical}) simply means that the interaction of spins
has to overcome the thermalization in order to produce significant spin
alignment. The temperature $T_{c}=\frac{J}{k}$ is usually called the
critical temperature and spontaneous symmetry breaking is possible only for $%
T<T_{c}.$ In this simple example, the non-zero magnetization $S$ breaks the
symmetry, $s_{i}\rightarrow -s_{i}.$ and originates from the competition
between spin-spin interaction which aligns the spin and the thermalization
which tends to destroy the alignment.\newline
Remarks:$\left( 1\right) $From the partition function in Eq(\ref{partition})
we see that the probability to find the system to have $x_{i}$ is given by
the Boltzmann factor 
\begin{equation}
P\left( x_{i}\right) =\frac{\exp (-\beta Nf\left( x_{i}\right) )}{%
\stackunder{i}{\sum }\exp (-\beta Nf\left( x_{i}\right) )}
\end{equation}
If $x_{0}$ is the absolute minimum for $f\left( x\right) $ , then in the
thermodynamic limit $N\rightarrow \infty ,$ we have $P\left( x_{0}\right)
\rightarrow 1$ and the probability for all the other $x_{i}$ will be zero. 
\newline
$\left( 2\right) $For the case $T$ is near $T_{c\text{ }},$the minimum of $%
f\left( x\right) $ is located at small values of $x$. Thus we can expand $%
f\left( x\right) $ in power series, 
\begin{equation}
f\left( x\right) =\frac{1}{2\beta _{c}}\left( 1-\frac{\beta }{\beta _{c}}%
\right) x^{2}+\frac{1}{12}\frac{\beta ^{3}}{\beta _{c}^{4}}x^{4}+\cdots
\end{equation}
The minimum is then 
\begin{equation}
x_{0}=\sqrt{\frac{3\beta _{c}^{3}}{\beta ^{3}}\left( 1-\frac{\beta }{\beta
_{c}}\right) }=\sqrt{\frac{3T^{3}}{T_{c}^{3}}\left( 1-\frac{T_{c}}{T}\right) 
}
\end{equation}
This means that near the critical temperature $T\rightarrow T_{0},$the
dependence of average magnetization on $\left( T-T_{0}\right) $ is
non-analytic. This is a typical behavior of physical quantities near the
critical point.

\subsection{Superfluid}

The superfluid $He^{4}$ provides a simple example of Goldstone excitation
where the excitation energy, $\varepsilon \left( k\right) $ goes to zero
when the wave number $k\rightarrow 0.$ The helium atoms are tightly bounded
and the long-distance attractive force between atoms are very weak while the
short distance is strongly repulsive. Thus a system of helium atoms can be
described as a gas of weakly interacting bosons with Hamiltonian, 
\begin{equation}
H=-\frac{1}{2m}\int d^{3}x\psi ^{\dagger }\nabla ^{2}\psi +\frac{1}{2}\int
d^{3}xd^{3}y\psi ^{\dagger }\left( x\right) \psi ^{\dagger }\left( y\right)
v\left( x-y\right) \psi \left( x\right) \psi \left( y\right)
\end{equation}
Here $v\left( x\right) $ is the potential describes the effective
interaction between helium atoms and $\psi \left( x\right) $ is the field
operator for the helium atom and satisfies the commutation relation, 
\[
\left[ \psi \left( x\right) ,\psi ^{\dagger }\left( y\right) \right] =\delta
^{3}\left( x-y\right) 
\]
We will assume that the system is in a large box of volume $\Omega $ with
periodical boundary condition. Clearly, this Hamiltonian is invariant under
the transformation, 
\begin{equation}
\psi \left( x\right) \rightarrow \psi ^{\prime }\left( x\right) =e^{i\alpha
}\psi \left( x\right)
\end{equation}
This is just a $U\left( 1\right) $ symmetry which says that the number of He
atoms is conserved. The conserved charge is just the number operator, 
\begin{equation}
Q=\int d^{3}x\psi ^{\dagger }\left( x\right) \psi \left( x\right)
\end{equation}
with the commutation relations, 
\begin{equation}
\left[ Q,\psi \left( x\right) \right] =\psi \left( x\right) ,\qquad \left[
Q,\psi ^{\dagger }\left( x\right) \right] =-\psi ^{\dagger }\left( x\right) .
\label{commutator}
\end{equation}
We can expand $\psi \left( x\right) $ in plane waves, 
\begin{equation}
\psi \left( x\right) =\frac{1}{\sqrt{\Omega }}\stackunder{\stackrel{%
\rightarrow }{k}}{\sum }a_{\stackrel{\rightarrow }{k}}e^{i\stackrel{%
\rightarrow }{k}\cdot \stackrel{\rightarrow }{x}}
\end{equation}
where $a_{\stackrel{\rightarrow }{k}}$ and $a_{\stackrel{\rightarrow }{k}%
}^{\dagger }$ are the usual creation and annihilation operators satisfying
the commutation relations, 
\begin{equation}
\left[ a_{\stackrel{\rightarrow }{k}},a_{\stackrel{\rightarrow }{k^{\prime }}%
}\right] =0,\left[ a_{\stackrel{\rightarrow }{k}},a_{\stackrel{\rightarrow }{%
k^{\prime }}}^{\dagger }\right] =\delta _{_{\stackrel{\rightarrow }{k}},_{%
\stackrel{\rightarrow }{k^{\prime }}}}
\end{equation}
The Hamilton is then of the form, 
\begin{equation}
H=\stackunder{k}{\sum }\frac{\hslash ^{2}k^{2}}{2m}a_{k}^{\dagger }a_{k}+%
\frac{1}{2\Omega }\stackunder{k_{i}}{\sum }\stackrel{\_}{v}\left(
k_{1}-k_{3}\right) \delta _{k_{1}+k_{2},k_{3}+k_{4}}a_{k_{1}}^{\dagger
}a_{k_{2}}^{\dagger }a_{k_{3}}a_{k_{4}}
\end{equation}
where 
\begin{equation}
\stackrel{\_}{v}\left( k\right) =\int d^{3}xe^{i\stackrel{\rightarrow }{k}%
\cdot \stackrel{\rightarrow }{x}}v\left( x\right)
\end{equation}
In these two equations and there after we have, for notational simplicity,
neglected the vector symbol for the wave vectors $k_{i}^{^{\prime }}$ s.
Since $v\left( x\right) $ is real we have $\stackrel{\_}{v}\left( k\right) =%
\stackrel{\_}{v}\left( -k\right) .$ For the trivial case where there is no
interaction, $v\left( x\right) =0,$ the ground state is just the one in
which all particles are in the $\stackrel{\rightarrow }{k}=0$ state, 
\begin{equation}
|\Psi _{0}>_{v=0}=\frac{\left( a_{0}^{\dagger }\right) ^{N}}{\sqrt{N!}}%
|0>\qquad \text{where \qquad }a_{k}|0>=0\text{ \quad }\forall k
\end{equation}
It is clear that if the interaction is small enough, in the ground state and
low-lying excited states, most of the particles will be in the $\stackrel{%
\rightarrow }{k}=0$ state , i.e. 
\begin{equation}
<n_{0}>\gg <n_{k}>\qquad \text{with \qquad }k\neq 0.
\end{equation}
where $n_{k}=a_{k}^{\dagger }a_{k}.$ We are interested in the cases where $%
N, $ the total number of particles, is very large. Thus $n_{0}\symbol{126}N$
is very large. From the properties of the creation and annihilation
operators, 
\begin{equation}
a_{0}|n_{0>}=\sqrt{n_{0}}|n_{0}-1>,\qquad a_{0}^{\dagger }|n_{0>}=\sqrt{%
n_{0}+1}|n_{0}+1>
\end{equation}
we will make the assumption that the matrix elements of $a_{0}$ are of order 
$\sqrt{n_{0}}$ and $a_{0}^{\dagger }$ of order $\sqrt{n_{0}+1}.$ Thus in the
limit $n_{0}\symbol{126}N\rightarrow \infty ,$ commutator of $a_{0\text{ }}$%
and $a_{0}^{\dagger },$ is of order unity while $a_{0},a_{0}^{\dagger }$ are
of order $\sqrt{N},$%
\begin{equation}
\left[ a_{0},a_{0}^{\dagger }\right] =1\ll a_{0}\,\quad \text{or\quad }%
a_{0}^{\dagger }\symbol{126}\sqrt{n_{0}}
\end{equation}
Thus we can neglect the commutator $.$ Since $a_{0},$ and $a_{0}^{\dagger }$
commute with all the other operators, 
\[
\left[ a_{0},a_{k}\right] =\left[ a_{0}^{\dagger },a_{k}\right] =0\qquad 
\text{for }k\neq 0. 
\]
We can then take $a_{0}$ and $a_{0}^{\dagger }$ to be c-numbers (Schur's
lemma), 
\begin{equation}
a_{0}=a_{0}^{\dagger }=\sqrt{n_{0}}
\end{equation}
Thus we will replace $a_{0}$ and $a_{0}^{\dagger }$ by $\sqrt{n_{0}}.$ Then
the coefficients of terms quadratic in $a_{k}$ and $a_{k}^{\dagger },$ $%
k\neq 0,$ in the interaction will be of order $n_{0}$and those of quartic
term is of order 1. Therefore we can make the approximation that neglect the
quartic terms and get the Hamiltonian in the form, 
\begin{eqnarray}
H &=&\stackunder{k\neq 0}{\sum }\frac{\hslash ^{2}k^{2}}{2m}a_{k}^{\dagger
}a_{k}+\frac{n_{0}}{2\Omega }\stackunder{k\neq 0}{\sum }[\stackrel{\_}{v}%
\left( k\right) (a_{k}^{\dagger }a_{-k}^{\dagger }+a_{k}a_{-k}) \\
&&+2\stackrel{\_}{v}\left( 0\right) a_{k}^{\dagger }a_{k}+2\stackrel{\_}{v}%
\left( k\right) a_{k}^{\dagger }a_{k}]+\frac{n_{0}^{2}}{2\Omega }\stackrel{\_%
}{v}\left( 0\right)
\end{eqnarray}
or 
\begin{equation}
H=\stackunder{k\neq 0}{\sum }\omega _{k}a_{k}^{\dagger }a_{k}+\frac{n_{0}}{%
2\Omega }\stackunder{k\neq 0}{\sum }\stackrel{\_}{v}\left( k\right)
(a_{k}^{\dagger }a_{-k}^{\dagger }+a_{k}a_{-k})+\frac{N^{2}}{2\Omega }%
\stackrel{\_}{v}\left( 0\right)  \label{pair H}
\end{equation}
where 
\begin{equation}
\omega _{k}=\frac{\hslash ^{2}k^{2}}{2m}+\frac{n_{0}}{\Omega }\stackrel{\_}{v%
}\left( k\right)
\end{equation}
and we have used 
\begin{equation}
N^{2}=\left( n_{0}+\stackunder{k\neq 0}{\sum }a_{k}^{\dagger }a_{k}\right)
^{2}\approx n_{0}^{2}+2n_{0}\stackunder{k\neq 0}{\sum }a_{k}^{\dagger }a_{k}
\end{equation}
Note that this Hamiltonian does not conserve the particle number but it
conserves the momentum because the removal of $k=0$ mode effects the
particle number but not the momentum. Since this Hamiltonian contains only
quadratic terms, we can solve this by Bogoliubove transformation as follows.
Define the quasi-particle operators by 
\begin{equation}
\alpha _{k}=\cosh \theta _{k}\text{ }a_{k}+\sinh \theta _{k}\text{ }%
a_{-k}^{\dagger }  \label{bogo1}
\end{equation}
then we have 
\begin{equation}
\alpha _{k}^{\dagger }=\cosh \theta _{k}\text{ }a_{k}^{\dagger }+\sinh
\theta _{k}\text{ }a_{-k}  \label{bogo2}
\end{equation}
where $\theta _{k}~$is an arbitrary parameter at our disposal. We now write
the Hamiltonian in terms of the quasi particle operators by inverting the
relations in Eq(\ref{bogo1,bogo2}), 
\begin{equation}
a_{k}=\cosh \theta _{k}\text{ }\alpha _{k}-\sinh \theta _{k}\text{ }\alpha
_{-k}^{\dagger },\qquad a_{-k}^{\dagger }=-\sinh \theta _{k}\text{ }\alpha
_{k}+\cosh \theta _{k}\text{ }\alpha _{-k}^{\dagger }
\end{equation}
and choose the parameter $\theta _{k}$ so that the coefficient of the
non-diagonal terms $\left( \alpha _{k}^{\dagger }\alpha _{-k}^{\dagger
}+\alpha _{k}\alpha _{-k}\right) $ is zero. The computation is
straightforward and the result is 
\begin{equation}
\tanh 2\theta _{k}=\frac{\frac{n_{0}\stackrel{\_}{v}}{\Omega }}{\omega _{k}}
\end{equation}
and the Hamiltonian is 
\begin{equation}
H=\stackunder{k\neq 0}{\sum }\varepsilon _{k}\alpha _{k}^{\dagger }\alpha
_{k}+\frac{N^{2}v\left( 0\right) }{2\Omega }+\frac{1}{2}\stackunder{k\neq 0}{%
\sum }\left( \varepsilon _{k}-\omega _{k}\right)
\end{equation}
where 
\begin{equation}
\varepsilon _{k}=\sqrt{\omega _{k}^{2}-\left( \frac{n_{0}\stackrel{\_}{v}}{%
\Omega }\right) ^{2}}=\sqrt{\left( \frac{\hslash ^{2}k^{2}}{2m}\right)
^{2}+2\left( \frac{\hslash ^{2}k^{2}}{2m}\right) \left( \frac{n_{0}\stackrel{%
\_}{v}\left( k\right) }{\Omega }\right) }
\end{equation}
This is just the Hamiltonian for the uncoupled harmonic oscillators and the
eigenvalues are 
\begin{equation}
E=\stackunder{k}{\sum }n_{k}\varepsilon _{k}
\end{equation}
The quai-particle energy excitation has the property that 
\begin{equation}
\varepsilon _{k}\rightarrow 0,\qquad \text{as \qquad }k\rightarrow 0
\end{equation}
which is just the Goldstone excitation. Clearly, the ground state $|\Psi
_{0}>$ is the one which is annihilated by all quasi particle operators $%
\alpha _{k,}$%
\begin{equation}
\alpha _{k}|\Psi _{0}>=0\qquad \forall \,k
\end{equation}
and the excited states are of the form, 
\begin{equation}
\left( \alpha _{k_{1}}^{\dagger }\right) ^{n_{1}}\left( \alpha
_{k_{2}}^{\dagger }\right) ^{n_{2}}\cdots |\Psi _{0}>
\end{equation}
It is straightforward to show that the quasi particle ground state can be
written in terms of the original creation operators $a_{k}^{\dagger }$ as 
\begin{equation}
|\Psi _{0}>=\sqrt{Z}\exp \{-\stackunder{k_{i}}{\sum }\tanh \,\theta
_{k_{i}}\,a_{k_{i}}^{\dagger }a_{-k_{i}}^{\dagger }\}|0>
\end{equation}
where 
\begin{equation}
Z=\stackunder{k_{i}}{\prod }\left( 1-\tanh ^{2}\theta _{k_{i}}\right)
\end{equation}
This shows that the quasi particle ground state is a complicate combination
of the vacuum, 2 particle states, 4 particle states,$\cdots $etc. To
elucidate the Goldstone theorem, we note that the vacuum expectation value
of $\psi \left( 0\right) $ in the ground state is non-zero, 
\begin{equation}
<\Psi _{0}|\psi \left( 0\right) |\Psi _{0}>=\frac{1}{\sqrt{\Omega }}<\Psi
_{0}|a_{0}|\Psi _{0}>=\sqrt{\frac{n_{0}}{\Omega }}\neq 0.
\end{equation}
where we have used the fact that $<\Psi _{0}|a_{k}|\Psi _{0}>=0,$ for $k\neq
0,$from the momentum conservation. From the commutation relation in Eq( \ref
{commutator}) we see that this is the symmetry breaking condition which
implies that 
\begin{equation}
Q|\Psi _{0}>\neq 0.
\end{equation}
The quasi particle excitation which has the property that its energy $%
\varepsilon _{k}$ goes to zero in the limit $k\rightarrow 0,$ is the
Goldstone excitation implies by the Goldstone's theorem.

\end{document}